\documentclass[twocolumn,english,reprint,english,jcp,aip,superscriptaddress]{revtex4-1}
\usepackage[T1]{fontenc}
\usepackage[latin9]{inputenc}
\usepackage{geometry}
\usepackage{array}
\usepackage{multirow}
\usepackage{amsmath}
\usepackage{graphicx}

\makeatletter

\providecommand{\tabularnewline}{\\}

\@ifundefined{textcolor}{}
{%
 \definecolor{BLACK}{gray}{0}
 \definecolor{WHITE}{gray}{1}
 \definecolor{RED}{rgb}{1,0,0}
 \definecolor{GREEN}{rgb}{0,1,0}
 \definecolor{BLUE}{rgb}{0,0,1}
 \definecolor{CYAN}{cmyk}{1,0,0,0}
 \definecolor{MAGENTA}{cmyk}{0,1,0,0}
 \definecolor{YELLOW}{cmyk}{0,0,1,0}
}

\makeatother

\usepackage{babel}
\begin{document}

\title{Embedded fragment stochastic density functional theory}

\author{Daniel Neuhauser}

\email{dxn@chem.ucla.edu}

\affiliation{Department of Chemistry, University of California at Los Angeles,
CA-90095 USA}

\author{Roi Baer}

\email{roi.baer@huji.ac.il}

\affiliation{Fritz Haber Center for Molecular Dynamics, Institute of Chemistry,
The Hebrew University of Jerusalem, Jerusalem 91904, Israel}

\author{Eran Rabani}

\email{eran.rabani@gmail.com}

\affiliation{School of Chemistry, The Sackler Faculty of Exact Sciences, Tel Aviv
University, Tel Aviv 69978, Israel}
\begin{abstract}
We develop a method in which the electronic densities of small fragments
determined by Kohn-Sham density functional theory (DFT) are embedded
using stochastic DFT to form the exact density of the full system.
The new method preserves the scaling and the simplicity of the stochastic
DFT but cures the slow convergence that occurs when weakly coupled
subsystems are treated. It overcomes the spurious charge fluctuations
that impair the applications of the original stochastic DFT approach.
We demonstrate the new approach on a fullerene dimer and on clusters
of water molecules and show that the density of states and the total
energy can be accurately described with a relatively small number
of stochastic orbitals. 
\end{abstract}
\maketitle
The desire to understand the structure and electronic properties of
complex hybrid materials and biological systems at the atomistic level
is the main motivation for developing fast large-scale electronic
structure approaches. One of the most successful theoretical frameworks
is density functional theory (DFT) within the Kohn-Sham (KS) formulation,\cite{Hohenberg1964,Kohn1965}
routinely used to model structures containing hundreds of electrons.\cite{kolb2012molecular,chelikowsky2011computational,frauenheim2002atomistic,Baroni2001,Siegbahn2009}
Formally, KS-DFT is thought to scale as $O\left(N^{3}\right)$, where
$N$ is the size of the system. This scaling prevents routine application
of KS-DFT to very large systems containing thousands of electrons
or more. While linear scaling techniques have been developed for KS-DFT,
their practical use is limited to low dimensional structures.\cite{Schwegler1996,Baer1997b,Goedecker1999,Scuseria1999,Soler2002,Skylaris2005,Gillan2007,Zeller2008,wang2008linear,Ozaki2010,rudberg2010kohn} 

Recently, we formulated KS-DFT as a statistical theory in which the
electron density is determined from an average of correlated stochastic
densities in a trace formula.\cite{Baer2013} As a result of self-averaging,
this so called stochastic DFT (sDFT) scales sub-linearly $O(N^{\varepsilon}),$
with $\varepsilon\le1$ for calculating the total energy per electron.
By controlling the stochastic fluctuations, the band structure, forces,
and density and its moments can also be described within sDFT. This
was illustrated for a series of silicon nanocrystals (NCs) of varying
sizes. 

Here we develop an embedded fragment version of sDFT (labeled efsDFT),
combining features from both the stochastic and embedded density functional
theories.\cite{daw1984embedded,Wesolowski1993,svensson1996oniom,govind1999electronic,Lin2007,elliott2010partition,goodpaster2010exact}
The efsDFT approach reduces the computational effort by decreasing
the number of stochastic orbitals required to converge the results
to a desired tolerance and at the same time circumvents a pathological
fault of sDFT associated with statistical noise caused by charge fluctuations
between weakly coupled fragments. The efsDFT approach is illustrated
for clusters of water molecules and for a fullerene dimer, two very
different test cases for which sDFT fails to provide an accurate estimate
of the electronic structure with a reasonable number of stochastic
orbitals, and as a result convergence of the self-consistent iterations
becomes sluggish. 

We first overview the derivation of sDFT. The starting point is the
expression of the total density of the full system, $n\left(\mathbf{r}\right)$,
as a trace:
\begin{equation}
n\left(\mathbf{r}\right)=tr\left\{ \hat{\theta}_{\beta}\hat{n}\left(\mathbf{r}\right)\right\} ,\label{eq:n}
\end{equation}
where $\hat{n}\left(\mathbf{r}\right)=\left|\mathbf{r}\left\rangle \right\langle \mathbf{r}\right|$
is the density operator and $\hat{\theta}_{\beta}=\mbox{erfc}\left(\beta\left(\mu-\hat{h}_{KS}\right)\right)$
is a smoothed representation of the density matrix. Here, $\beta$
is a smoothing inverse energy parameter chosen such that $\beta^{-1}\ll E_{g}$,
where $E_{g}$ is the HOMO-LUMO gap. Note that $\lim_{\beta\rightarrow\infty}\mbox{erfc}\left(\beta x\right)=2\theta\left(x\right)$,
where $\theta\left(x\right)$ is the Heaviside function and the factor
of ``$2$'' accounts for electron spin. In the above, $\hat{h}_{KS}$
is the KS Hamiltonian of the full system which depends on the full
density $n\left(\mathbf{r}\right)$. The chemical potential $\mu$
is determined by requiring that the density integrates to $N$ electrons. 

In sDFT we use the stochastic trace formula to evaluate Eq.~\eqref{eq:n}.
The procedure consists of: 
\begin{itemize}
\item Generating a set of $I$ stochastic orbitals $\chi\left(\mathbf{r}\right)$
on the grid. 
\item For each $\chi\left(\mathbf{r}\right)$, calculating the random-occupied
orbital $\zeta\left(\mathbf{r}\right)=\sqrt{\hat{\theta}_{\beta}}\chi\left(\mathbf{r}\right)$
($\sqrt{\hat{\theta}_{\beta}}$ operates on $\chi\left(\mathbf{r}\right)$
using a suitable expansion in terms of Chebyshev polynomials\cite{Kosloff1988}).
\item Averaging (symbolized by $\left\langle \cdots\right\rangle _{\chi}$)
over the square of the random occupied orbital gives an estimate of
the density:
\begin{equation}
n\left(\mathbf{r}\right)=\left\langle \left|\zeta\left(\mathbf{r}\right)\right|^{2}\right\rangle _{\chi}.\label{eq:FullStochastic}
\end{equation}
$\left|\zeta\left(\mathbf{r}\right)\right|^{2}$ is a random variable
distributed with mean $n\left(\mathbf{r}\right)$ given by the exact
non-interacting ground state density of $\hat{h}_{KS}$ at point $\mathbf{r}$
and with variance given by $\frac{\sigma_{0}\left(\mathbf{r}\right)}{\sqrt{I}}$,
where $\sigma_{0}\left(\mathbf{r}\right)$ is determined by the properties
of the underlying physical/chemical system. 
\end{itemize}
The control of the error is done through the number of stochastic
orbitals $I$. Any method to reduce $\sigma_{0}$ will allow a corresponding
reduction of $I$ therefore improving efficiency. One way to achieve
this is by limiting the stochastic average to a small difference between
the full and approximate density operator which will thus exhibit
a smaller $\sigma_{0}$ (for a similar use in a related field, Auxiliary
Field Monte Carlo, see\textit{ }\textit{\emph{Ref.}} \onlinecite{rom1997shifted}).
Such an approximate operator can be obtained from a division of the
system into $F$ small fragments, where each fragment $f=1,\dots,F$
has its own set of atomic cores and its own KS Hamiltonian, $\hat{h}_{KS}^{\left(f\right)}$.
The KS Hamiltonian of each fragment can be constructed from the external
potential of the atomic cores in each fragment. Each fragment $f$
is now assigned to have $N^{\left(f\right)}$ electrons such that
the total number of electrons is $\sum_{f}N^{\left(f\right)}=N$.
The density $n^{\left(f\right)}\left(\mathbf{r}\right)$ can be determined
separately for each fragment using KS-DFT. This produces occupied
and low-lying unoccupied KS eigenstates (indexed by $j$) $\varphi_{j}^{\left(f\right)}\left(\mathbf{r}\right)$
and eigenvalues $\varepsilon_{j}^{\left(f\right)}$. One can now write
an approximation to $n\left(\mathbf{r}\right)$ in terms of the sum
of fragmented densities as:

\begin{equation}
n\left(\mathbf{r}\right)\approx n_{F}\left(\mathbf{r}\right)=\sum_{f=1}^{F}n^{(f)}\left(\mathbf{r}\right),
\end{equation}
where the density $n^{(f)}\left(\mathbf{r}\right)$ in each fragment
can also be expressed as a trace, $n^{(f)}\left(\mathbf{r}\right)=tr\left\{ \hat{\theta}_{\beta}^{(f)}\hat{n}\left(\mathbf{r}\right)\right\} $
with
\begin{equation}
\hat{\theta}_{\beta}^{(f)}=\sum_{j}\mbox{erfc}\left(\beta\left(\mu^{(f)}-\varepsilon_{j}^{\left(f\right)}\right)\right)\left|\varphi_{j}^{\left(f\right)}\left\rangle \right\langle \varphi_{j}^{\left(f\right)}\right|.
\end{equation}
The stochastic trace in Eq.~\eqref{eq:FullStochastic} can therefore
be replaced by an embedding form: 
\begin{equation}
n\left(\mathbf{r}\right)=n_{F}\left(\mathbf{r}\right)+\left\langle \left|\zeta\left(\mathbf{r}\right)\right|^{2}-\sum_{f=1}^{F}\left|\zeta^{\left(f\right)}\left(\mathbf{r}\right)\right|^{2}\right\rangle _{\chi},\label{eq:StochFragDensity}
\end{equation}
where $\zeta^{\left(f\right)}\left(\mathbf{r}\right)=\sqrt{\hat{\theta}_{\beta}^{(f)}}\chi\left(\mathbf{r}\right)$.
The density obtained from Eq.~\eqref{eq:StochFragDensity} is used
to construct a new KS Hamiltonian $\hat{h}_{KS}$ and the procedure
is repeated and converged to the final self-consistent field (SCF)
solution using DIIS~\cite{Pulay1980} within typically less than
10 SCF iterations. The advantage of Eq.~\eqref{eq:StochFragDensity}
is clear: as $n_{F}\left(\mathbf{r}\right)\rightarrow n\left(\mathbf{r}\right)$
the variance $\sigma_{0}\left(\mathbf{r}\right)$ decreases, reducing
the number of stochastic orbitals required for convergence at a desired
tolerance. The use of $n_{F}\left(\mathbf{r}\right)$ dramatically
reduces spurious charge fluctuations between fragments induced by
poor statistical sampling in the original sDFT approach. Because of
such fluctuations sDFT requires a large number of stochastic orbitals
for convergence while esDFT, which does not suffer from the spurious
fluctuations, requires only few tens or hundreds of stochastic orbitals.
Further, as long as each fragment is not too large there is very little
additional computational overhead and the scaling of the method is
unchanged.

\begin{figure}[t]
\begin{centering}
\includegraphics[width=8cm]{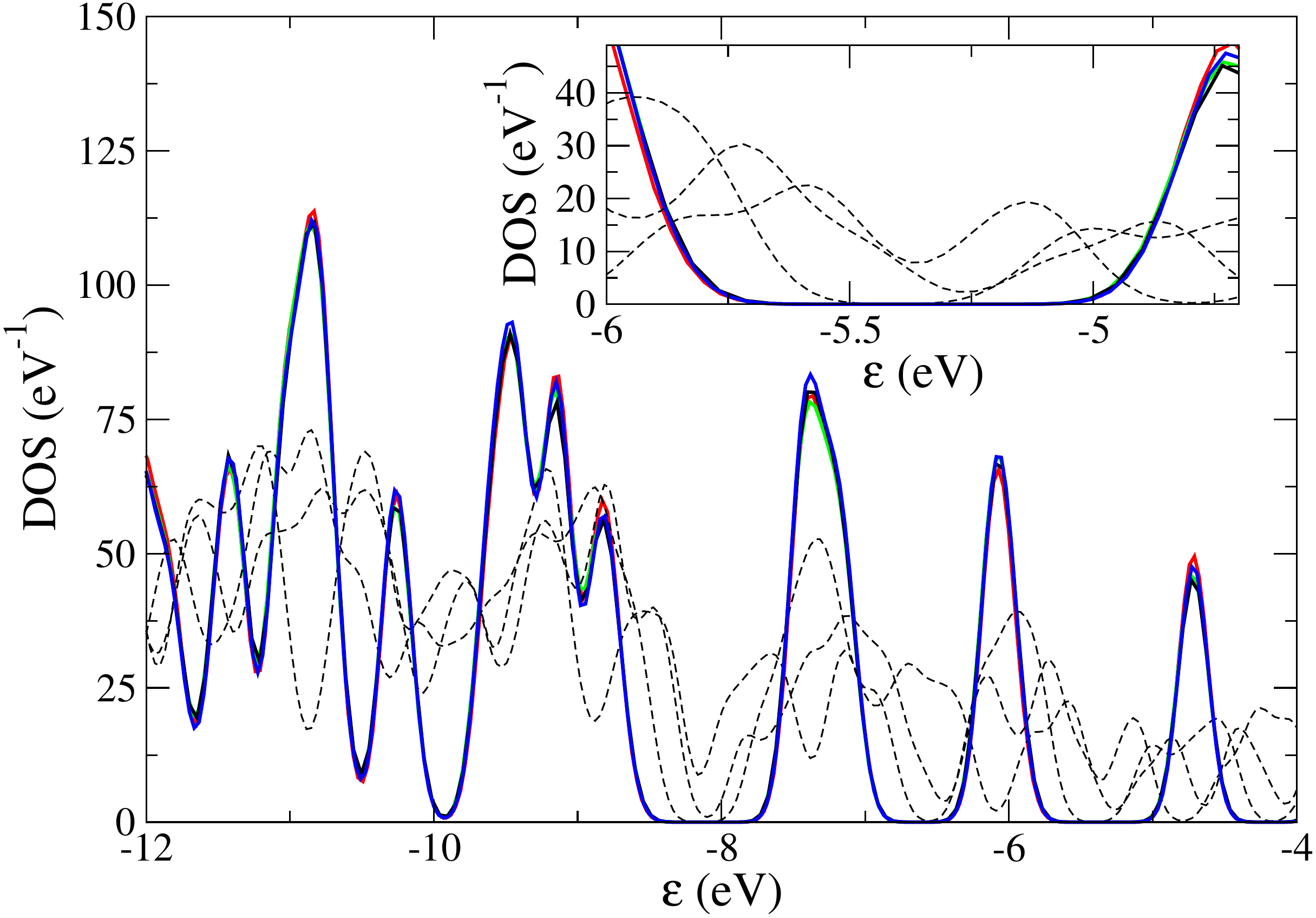}
\par\end{centering}

\protect\caption{\label{fig:C60Dimer-DOS}The density of states of a fullerene dimer
calculated using a deterministic DFT approach (black curve), efsDFT
with $I=80$ (red curve), $I=160$ (green curve) and $I=320$ (blue
curve) stochastic orbitals. The dashed curves are sDFT results with
$I=320$ stochastic orbitals for three different initial seeds. Inset:
zoom into the energetic region of the gap. }
\end{figure}

We tested two generic cases for efsDFT and compared the results with
sDFT and with a deterministic DFT approach (free of statistical errors),
labeled dDFT below. The first test case involves a fullerene dimer
with center-to-center separation of $\approx1\mbox{nm}$ (the equilibrium
value of bulk fullerene) as shown in Fig.~\ref{fig:C60Dimer-DOS}.
At this separation, the perturbation in the charge density of each
fullerene caused by the neighboring monomer is rather small. Results
based on sDFT using $I=320$ stochastic orbitals are shown for three
different seeds (dashed curves). We find significant deviations of
the density of states (DOS), caused by fictitious charge transfer
between the monomers, and equally striking is the spread of the results.
The charge sloshing phenomenon appears because of stochastic fluctuations,
which in the case of weak coupling between the fragments leads to
a spurious finite density of states inside the HOMO-LUMO gap. Increasing
the number of stochastic orbitals will eventually fix this problem
but at a much higher numerical cost. In fact, the number of stochastic
orbitals required to converge the results in sDFT increases for weaker
coupling between the fragments. 

\begin{figure}[b]
\begin{centering}
\includegraphics[width=8cm]{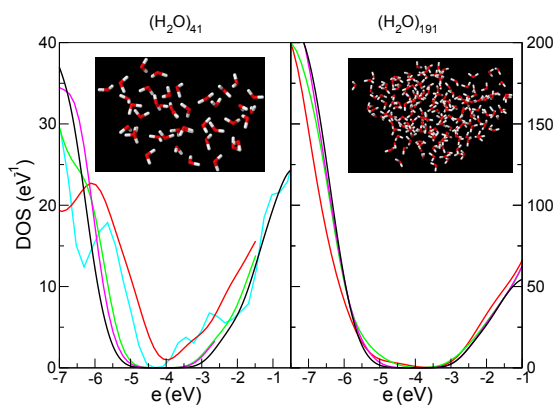}
\par\end{centering}

\protect\caption{\label{fig:water}The density of states near the highest occupied
and lowest unoccupied KS eigenvalues of $\mbox{\ensuremath{\left(H_{2}O\right)_{41}}}$
(left panel) and $\mbox{\ensuremath{\left(H_{2}O\right)_{191}}}$
(right panel) using sDFT with $I=320$ (cyan) and efsDFT with $I=80$
(red), $160$ (green) and $320$ (magenta) stochastic orbitals. The
solid black curve represents the deterministic DFT calculation. }
\end{figure}

\begin{table*}[t]
\protect\caption{\label{tab:SEConvergence}efsDFT (and in one case sDFT) based results
and corresponding deterministic values (dDFT) for the three systems
studied for different number of random orbitals $I$. The following
energies (in eV) are considered: HOMO ($\varepsilon_{HOMO}$) and
LUMO ($\varepsilon_{LUMO}$) , total energy $E_{tot}$ and total energy
per electron $E_{tot}/N_{e}$. The numbers in parenthesis are the
standard deviation in the last given digit(s) estimated from 5 independent
runs.}

\centering{}%
\begin{tabular}{|c|c|c|r@{\extracolsep{0pt}.}l|c|c|c|}
\hline 
System & Method & $I$ & \multicolumn{2}{c|}{$\varepsilon_{HOMO}$} & $\varepsilon_{LUMO}$ & $E_{tot}$ & $E_{tot}/N_{e}$\tabularnewline
\hline 
\hline 
\multirow{4}{*}{$\left(\mbox{\ensuremath{H_{2}}O}\right)_{41}$} & \multirow{3}{*}{efsDFT} & $80$ & \multicolumn{2}{c|}{$-5.8\left(3\right)$} & $-2.5\,\left(2\right)$ & $-19129\,\left(1\right)$ & $-58.320\,\left(4\right)$\tabularnewline
\cline{3-8} 
 &  & $160$ & \multicolumn{2}{c|}{$-6.1\,\left(1\right)$} & $-2.3\,\left(1\right)$ & $-19128\,\left(1\right)$ & $-58.317\,\left(4\right)$\tabularnewline
\cline{3-8} 
 &  & $320$ & \multicolumn{2}{c|}{$-6.1\,\left(1\right)$} & $-2.3\,\left(1\right)$ & $-19126.9\,\left(7\right)$ & $-58.314\,\left(2\right)$\tabularnewline
\cline{2-8} 
 & \multirow{1}{*}{dDFT} &  & \multicolumn{2}{c|}{$-5.9$} & $-2.4$ & $-19127.0$ & $-58.314$\tabularnewline
\hline 
\multirow{4}{*}{$\left(\mbox{\ensuremath{H_{2}}O}\right)_{191}$} & \multirow{3}{*}{efsDFT} & $80$ & \multicolumn{2}{c|}{$-5.0\,\left(3\right)$} & $-2.60\,\left(5\right)$ & $-89212\,\left(3\right)$ & $-58.385\,\left(2\right)$\tabularnewline
\cline{3-8} 
 &  & $160$ & \multicolumn{2}{c|}{$-5.5\,\left(2\right)$} & $-2.66\,\left(7\right)$ & $-89210\,\left(1\right)$ & $-58.384\,\left(1\right)$\tabularnewline
\cline{3-8} 
 &  & $320$ & \multicolumn{2}{c|}{$-5.7\,\left(1\right)$} & $-2.55\,\left(6\right)$ & $-89209\,\left(1\right)$ & $-58.383\,\left(1\right)$\tabularnewline
\cline{2-8} 
 & \multirow{1}{*}{dDFT} &  & \multicolumn{2}{c|}{$-5.6$} & $-2.48$ & $-89208$ & $-58.382$\tabularnewline
\hline 
\multirow{5}{*}{$\mbox{\ensuremath{C_{60}}-\ensuremath{C_{60}}}$} & sDFT & $320$ & \multicolumn{2}{c|}{$-5.559\,\left(130\right)$} & $-4.889(122)$ & $-18701.0\,\left(30\right)$ & $-38.9610\,\left(60\right)$\tabularnewline
\cline{2-8} 
 & \multirow{3}{*}{efsDFT} & $80$ & \multicolumn{2}{c|}{$-5.925\,\left(25\right)$} & $-4.823\,\left(29\right)$ & $-18713.3\,\left(5\right)$ & $-38.9861\,\left(10\right)$\tabularnewline
\cline{3-8} 
 &  & $160$ & \multicolumn{2}{c|}{$-5.964\,\left(9\right)$} & $-4.755\,\left(21\right)$ & $-18713.1\,\left(3\right)$ & $-38.9857\,\left(6\right)$\tabularnewline
\cline{3-8} 
 &  & $320$ & \multicolumn{2}{c|}{$-5.969\,\left(2\right)$} & $-4.752\,\left(4\right)$ & $-18713.3\,\left(2\right)$ & $-38.9861\,\left(4\right)$\tabularnewline
\cline{2-8} 
 & \multirow{1}{*}{dDFT} &  & \multicolumn{2}{c|}{$-5.973$} & $-4.746$ & $-18713.1$ & $-38.9857$\tabularnewline
\hline 
\end{tabular}
\end{table*}

Using the efsDFT with deterministic KS orbitals taken from each of
the monomers yields a very rapid convergence of the DOS with the number
of stochastic orbitals, as shown in Fig.~\ref{fig:C60Dimer-DOS}
(red, green and blue curves). Importantly, a clear HOMO-LUMO gap is
observed even when we use a very small number of stochastic orbitals.
Furthermore, we do not observe the aforementioned spurious charge
transfer between the two monomers. 

The second test case involves two water clusters, with $41$ and $191$
molecules. The purpose is (1) to study a system with short-range order
and long-range disorder and (2) to explore the efsDFT computational
scaling with system-size. We used molecular dynamics (MD) with the
flexible SPC forces field and a smooth cutoff~\cite{fennell2006ewald}
to generate the disordered structures. The last time step configuration
of the equilibrated trajectory was taken as the input structure for
the efsDFT, sDFT and dDFT calculations.

In Fig.~\ref{fig:water} we compare the efsDFT and dDFT calculations
for the two water clusters. The sDFT calculations, which are shown
only for the smaller cluster with $I=320$, preserve a gap in the
density of states near the Fermi energy. However, due to unrealistic
charge fluctuations there is a pronounced shift in the Fermi energy
and a significant deformation of the DOS. In contrast, the efsDFT
calculations, which used individual water molecules as the fragments,
displays quantitative DOS already for $I=160$. 

In Table~\ref{tab:SEConvergence} we summarize the results for the
HOMO and LUMO orbital energies, and the total energy and total energy
per electron. For the fullerene dimer, at $I=320$, sDFT deviates
from the deterministic approach by $\approx400\mbox{meV}$ and $\approx150\mbox{meV}$
for the HOMO and LUMO orbital energies, respectively, while efsDFT
is accurate to within a few meV's. Moreover, the total energy per
electron in sDFT deviates significantly from the deterministic value
while efsDFT provides an accurate estimate to within a fraction of
an meV. 

A similar picture emerges for the water clusters. For example, the
statistical error and the deviation from the deterministic approach
in the HOMO and LUMO orbital energies are $50-100\,\mbox{meV}$ using
$I=320$ for the larger water cluster. Further, the statistical error
and deviation from deterministic values of the orbital and per-electron
energies decrease with cluster size for a fixed number of stochastic
orbitals, indicating self-averaging.\cite{Baer2013} Since the scaling
of the approach with system size is \emph{linear} for a fixed number
of stochastic orbitals, this self-averaging suggests that for a given
statistical error the new approach scales \emph{sub-linearly}, similar
to sDFT for homogeneous covalent systems. 

In summary, we presented a new DFT method which combines features
from both embedded and stochastic density functional theories. The
densities of small fragments of the system were calculated by a deterministic
DFT approach and were used to reconstruct the total density of the
system deploying stochastic orbitals in a trace formula. The resulting
method, so called efsDFT, preserves the scaling of sDFT, including
the concept of self-averaging. Moreover, it overcomes some the limitations
of sDFT, specifically for weakly coupled systems, achieving much faster
convergence with the number of stochastic orbitals for both the density
of states as well as for the total energy of the system. This was
shown for two generic models, a weakly bound fullerene dimer and disordered
clusters of water molecules.

efsDFT could be improved by a more sophisticated choice of the fragments,
i.e., one that minimizes the density difference $\left|n\left(\mathbf{r}\right)-n_{F}\left(\mathbf{r}\right)\right|$.
For example, one could self consistently improve the fragment Hamiltonians
during the SCF iterations or ``carve'' them out of the total potential
if easier. Even overlapping fragments could be used. This is because
Eq. \ref{eq:StochFragDensity} is exact regardless of the choice of
$n_{F}\left(\mathbf{r}\right)$. Work along these lines and others
is currently in progress.

R. B. and E. R. gratefully thank the Israel Science Foundation, Grants
No. 1020/10 and No. 611/11, respectively. R. B. and D. N. acknowledge
the support of the US-Israel Bi-National Science Foundation. D. N.
gratefully acknowledges support by the NSF, grant CHE-1112500.

\end{document}